\documentclass[preprint,review,sort&compress,12pt]{elsarticle}
\usepackage[english,russian]{babel}
\usepackage{hyperref}
\hypersetup{
    colorlinks=true,        
    linkcolor=blue,          
    citecolor=blue,        
    filecolor=magenta,      
    urlcolor=blue           
}

\usepackage{amssymb}

\usepackage{amsmath}

\usepackage[modulo]{lineno}

\journal{Nuclear Instruments and Methods in Physics Research A}

\begin{document}

\begin{frontmatter}



\title{Acceleration and Focusing Electron/Positron Bunches in Plasma-Dielectric Wakefield Accelerator}


\author[1]{Gennadiy V. Sotnikov\corref{cor1}}
\ead{sotnikov@kipt.kharkov.ua}
\cortext[cor1]{Corresponding author}

\author[1]{Kostyantyn V. Galaydych}
\ead{kgalaydych@gmail.com}

\author[2]{Jay L. Hirshfield}
\ead{jay.hirshfield@yale.edu}

\author[1]{Peter I. Markov}
\ead{pmarkov11@outlook.com}

\author[1]{Ivan M. Onishchenko}
\ead{onish@kipt.kharkov.ua}

\affiliation[1]{organization={National Science Center Kharkov Institute of Physics and Technology},
            addressline={1, Akademicheskaya St.},
            city={Kharkiv},
            postcode={61108},
            country={Ukraine}
}

\affiliation[2]{organization={Department of Physics, Yale University},
            city={New Haven, Connecticut},
            postcode={O6511},
            country={USA}
}


\begin{abstract}
To mitigate the BBU instability and improve characteristics of accelerated bunches in Dielectric Wakefield Accelerator one can be used the isotropic plasma filling of the transport channel.
Here we present the results of analytical and numerical studies of the dynamics of accelerated electron/positron and drive electron bunches under wake acceleration in a plasma DWA (PDWA)  with a vacuum channel. The wake field is excited by an electron bunch in a quartz dielectric tube inserted into a cylindrical metal waveguide. The inner region of the dielectric tube is filled with plasma with a vacuum channel along the waveguide axis. At the numerical simulations the energy and spatial characteristics, efficiency, emittance, and energy spread for accelerated positron and electron bunches is studied for different radii of the vacuum channel.

The transverse instability of the drive bunch in PDWA is studied analytically and numerically. The analytical studies have discovered the presence of one surface and one bulk eigenwaves, which are absent in corresponding dielectric-loaded waveguide without plasma filling. The main contribution to amplitude of transverse wakefield, responsible for stabilization of transverse motion of bunches, brings the bulk plasma eigenwave. The comparative analysis of the data resulting from analytical studies and the ones obtained by numerical simulation has demonstrated qualitative agreement between the results.
\end{abstract}




\end{frontmatter}


\section{Introduction}\label{sec:Intro}
Due to their compactness, dielectric wakefield accelerators (DWA), which are one of the varieties of structure-based wakefield accelerators (SWFA)~\cite{Jing_2022,lu2022advancedrfstructureswakefield}, can find application in various fields of industry, technology, medicine, etc. They are also currently considered as one of the promising candidates for building linear colliders~\cite{vieira2024reportadvancedlinearcollider}. The ease of manufacturing the slowing down structure, the stability of the wake wave, as well as the invariance with respect to the charge of the accelerated bunch make dielectric wakefield accelerators attractive for use in collider purposes. However, charged bunches of particles in a dielectric wakefield accelerator, as in any linear accelerator, are subject to beam breakup instability. This limits the acceleration rate~\cite{Li2014PRSTAB, Baturin2017PRSTAB} and degrades the quality of the accelerated bunches, which is especially relevant for collider applications. To improve bunch transport in dielectric wake accelerators, it was proposed to fill the vacuum channel with plasma~\cite{Sot2014NIMA}. Subsequent analytical studies and PIC numerical simulations showed that plasma-filled dielectric wake accelerators (PDWA) make it possible to focus accelerated electron bunches~\cite{Markov2016PAST,BIAGIONI2018247,Sot2020JoI-9} and suppress the instability of the drive bunch~\cite{GALAYDYCH2022166766}.

It would seem that the presence of plasma in the DWA transport channel could worsen (or even disrupt) the transport of the accelerated positron bunch, as it happens in beam driven plasma wakefield accelerators (PWFA). Various methods have been considered to improve the quality of accelerated positron bunches in PWFA. One of them is to create a on-axis vacuum channel inside the plasma region~\cite{Ges2016Ncom}. Our numerical simulations show that the presence of a vacuum channel in PDWA with different transverse plasma density profiles (both homogeneous plasma and capillary discharge plasma~\cite{Bob2001PRE,Stein2006PRSTAB}) can even improve the focusing of the accelerated positron bunch~\cite{Markov_2022} compared to completely filling the dielectric tube. The physical reason for the presence of focusing of the positron bunch is the formation of a paraxial region with an excess of negative charge due to the returned plasma electrons. The formation of such a region is similar to that formed when the plasma column is blown by a drive bunch in one of the PWFA schemes~\cite{Diederichs_PRAB2019}. The presence of a vacuum channel in the plasma does not suppress the beam breakup instability of the drive bunch in PWFA~\cite{Schroeder1999PRL}. But in PDWA, the instability of the drive bunch does not develop or it can be significantly diminished~\cite{GALAYDYCH2024169156}. The physical picture of the diminishing of the beam breakup instability in PDWA consists in the excitation of the focusing wave by the drive bunch along with the accelerating wave. The latter can be either a volume~\cite{GALAYDYCH2022166766} and/or a surface~\cite{GALAYDYCH2024169156} plasma wave.

As noted above, our numerical simulation confirmed the predictions of the analytical theory, demonstrating the acceleration of a test electron or positron bunch with its simultaneous focusing~\cite{Sot2020JoI-9,Markov_2022}. At these studies, the parameters of the plasma (density, size of the vacuum channel) and bunches (charge, size, delay time) in the case of electron acceleration differed from the case of positron acceleration. In addition, the characteristics of accelerated electron bunches were not studied at all. In this paper, we study the transportation of accelerated electron and positron bunches in a plasma-dielectric wake accelerator in the same problem formulation with the same parameters of the accelerator structure, drive bunch, and electron and positron bunch (except for their delay time). For the same parameters of the accelerator structure, using analytical and numerical methods (for linear and quasi-linear plasma cases, respectively), the stability of the off-axis drive bunch is investigated and a comparative analysis of the excited transverse wake field is carried out for the cases of the presence and absence plasma in the transport channel.

The article is organized as follows: in section~\ref{sec:2} we present the general statement of the problem and give parameters of the structure and bunches used in PIC numerical simulation. In section~\ref{sec:3} the results of the 2.5-dimensional PIC-simulation of positron acceleartion in PDWA are presented. In section~\ref{sec:4} the results analytical and  numerical  studies of BBU instability in PDWA is given. Section~\label{ref:05} concludes this work.

\section{Statement of the problem}\label{sec:2}
The waveguide structure of the investigated plasma-dielectric wakefield accelerator (PDWA) is a dielectric tube with an inner radius of $a$ and an outer radius of $b$, inserted into a cylindrical metal waveguide. The inner region of the dielectric tube between the radii of $r_{\rm v}$ and $a$ is filled with plasma. Thus, in the general case, there is a vacuum channel of radius $r_{\rm v}$ in the accelerating structure. A cylindrical drive electron bunch of radius $R_b$ is injected along the axis of the slow-wave structure through the input end closed by a conductive grid transparent for electrons. After a delay time of $t_{del}$ after the drive bunch injection, a test cylindrical electron or positron bunch of radius $r_w$ is injected through the input end, also along the axis. Radius $r_w$ can be larger than the radius of the vacuum channel. In subsequent simulation the absolute value of the test bunch charge was assumed to be much smaller than that of the drive bunch.

For numerical simulation of wake fields and dynamics of driver and test bunches we used our own 2.5-dimensional particle-in-cell (PIC) code~\cite{Sot2014NIMA}. The parameters of the structure and bunches used in the simulation are given in Table~\ref{Tabl:01}.
\begin{table}[h!]
\begin{center}
 \caption{Parameters of the structure and bunches used in the simulation.}
 \label{Tabl:01}
 \begin{tabular}{| l | r| }
  \hline \hline
  Outer dielectric-tube radius $b$ & 0.6~mm \\
  Inner dielectric-tube radiu $a$ & 0.5~mm \\
  Outer plasma-cylinder radius $b$ & 0.5~mm \\
  Inner plasma-cylinder radius (vacuum channel) $r_{\rm v}$ & $0\div0.5$~mm \\
  Waveguide length $L$ & 80~mm \\
  Dielectric permittivity $\varepsilon$ & 3.75 (quartz) \\
  Bunch energy $E_0$ & 5~GeV \\
  Drive electron bunch charge & $-3$~nC \\
  Witness electron/positron bunch charge & $\mp 0.05$~nC \\
  Longitudinal rms deviation of drive bunch charges $2\sigma_1$  & 0.1~mm \\
  Longitudinal rms deviation of test bunch charges $2\sigma_2$  & 0.05~mm \\
  Total drive bunch length used in PIC simulation & 0.2~mm \\
  Total test bunch length used in PIC simulation & 0.1~mm \\
  Drive bunch diameter $2R_b$ & 0.9~mm \\
  Electron/positron test bunch diameter $2r_w$ & 0.7~mm \\
  Plasma density & $2\cdot10^{14}\,\mbox{cm}^{-3}$ \\
  \hline \hline
  \end{tabular}
  \end{center}
  \end{table}

When the plasma simulating it was assumed that at the initial moment it has a homogeneous density. A homogeneous plasma profile with a hollow channel can be obtained using beams described by high-order Bessel functions~\cite{Ges2016Ncom,Kimura:2011zz}. To simulate the positron and electron bunches acceleration the same plasma density was used. This density differs from that used in our work~\cite{Sot2020JoI-9} and coincides with that given in~\cite{Markov_2022}. The choice of the same density facilitates the comparison of the focusing positrons and electrons capabilities, although it may not be optimal for each individual case.

\section{Numerical analysis of accelerated electron and positron bunches focusing}\label{sec:3}

In this section, the behavior of the transverse size of accelerated electron and positron bunches, as well as their emittance and energy spread during their transportation in PDWA are investigated.

\begin{figure}[!th]
  \centering
  \includegraphics[width=0.5\textwidth]{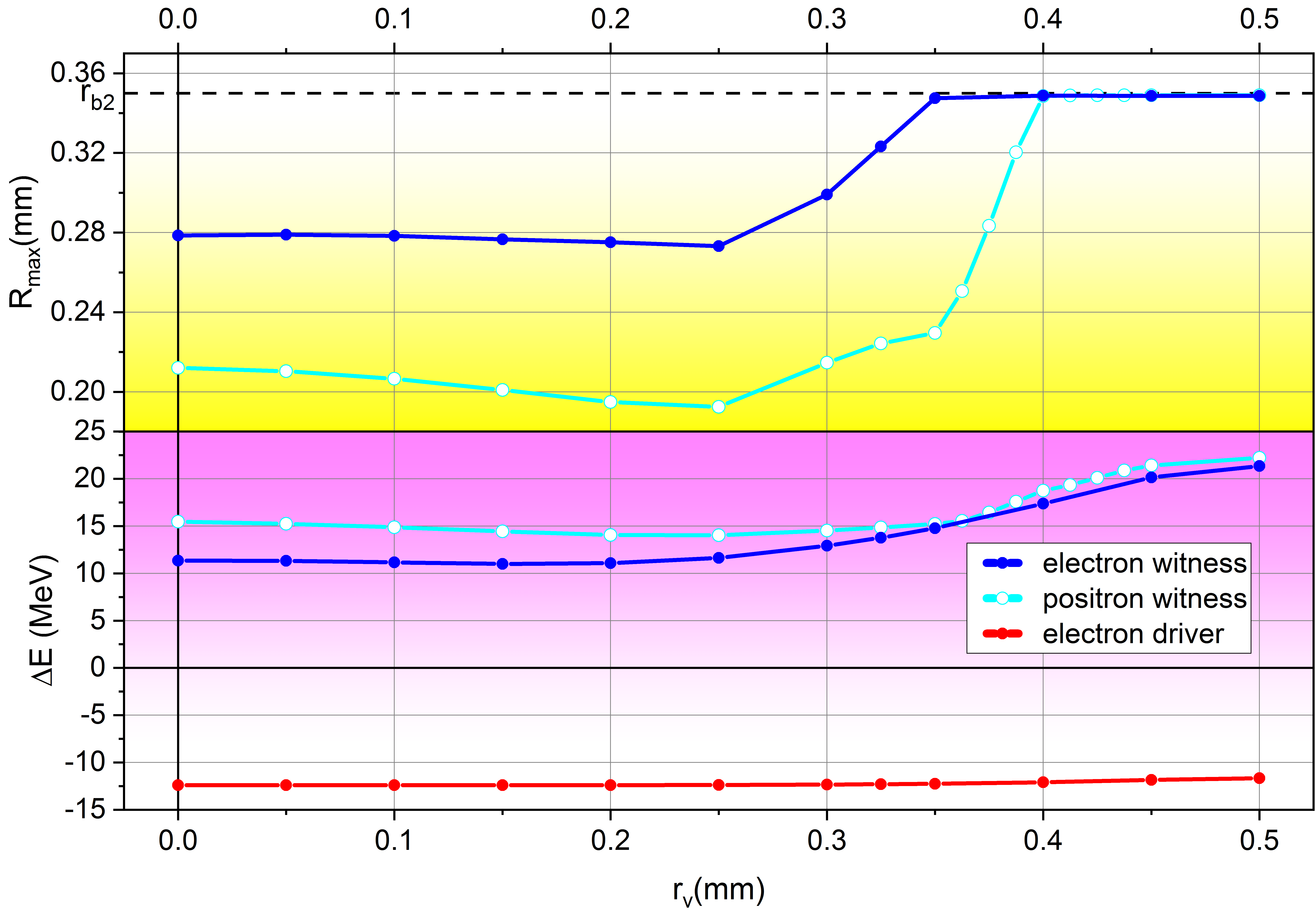}
  \caption{Behavior of the radius of the test bunch $R_{max}$ (top) and the energy gain $\Delta E$ of accelerated bunches of electrons, positrons and a slow drive electron bunch (bottom) with a change in the radius of the vacuum channel $r_{\rm v}$ for the time $t=266.9$~ps (the drive bunch reached the end of the waveguide).}\label{Fig:01}
\end{figure}

The behavior of the radius of the test electron and positron bunches $R_{max}$ with a change in the vacuum channel radius $r_{\rm v}$ from 0 to 0.5~mm for the time $t=266.9$~ps (at this moment the drive bunch reached the end of the waveguide) is shown in Fig.~\ref{Fig:01} in the upper part. Here $R_{max}$ means the radius of the outermost particle in the bunch. As follows from the curves shown in Fig.~\ref{Fig:01}, with an increase in $r_{\rm v}$ from 0 to 0.25~ mm, a gradual increase in the focusing of the test bunches is observed. With a subsequent increase in $r_{\rm v}$ to $r_{\rm v}=0.35$~mm, the focusing deteriorates somewhat. Further increase in $r_{\rm v}$ leads to a sharp drop in the focusing of the test bunches. At $r_{\rm v}\geq 0.425$~mm there is no focusing. As a result of focusing, the change in the diameter of the test electron bunch is approximately 2 times smaller than that of the positron bunch. This is due to the fact that for the same sizes of the electron and positron bunches that we used, at the selected delay times, the focusing transverse force $F_r$ acting on the electrons is approximately 1.7 times smaller than the force $F_r$ acting on the positrons.  Changing the initial parameters of the bunches, as well as changing the plasma density, can lead to the opposite situation, when the transverse focusing for the test electron bunch can be stronger than for the positron bunch.

The lower part of Fig.~\ref{Fig:01} shows the dependence of the energy gain $\Delta E$ of the test bunches (blue and light blue curves) and the slowed drive bunch (red curve) on the change in the vacuum channel size $r_{\rm v}$. With an increase in $r_{\rm v}$ from 0 to $0.15\div 0.25$~mm, a slight decrease in the energy gain of the accelerated bunches $\Delta E$ is observed for both electrons and positrons. A further increase in $r_{\rm v}$ from 0.25~mm to 0.5~mm leads to an increase in the energy gain of the accelerated bunches $\Delta E$. The energy of the drive bunch changes weakly over the acceleration length under consideration due to a significantly larger charge.

\begin{figure}[!th]
  \centering
  \includegraphics[width=\textwidth]{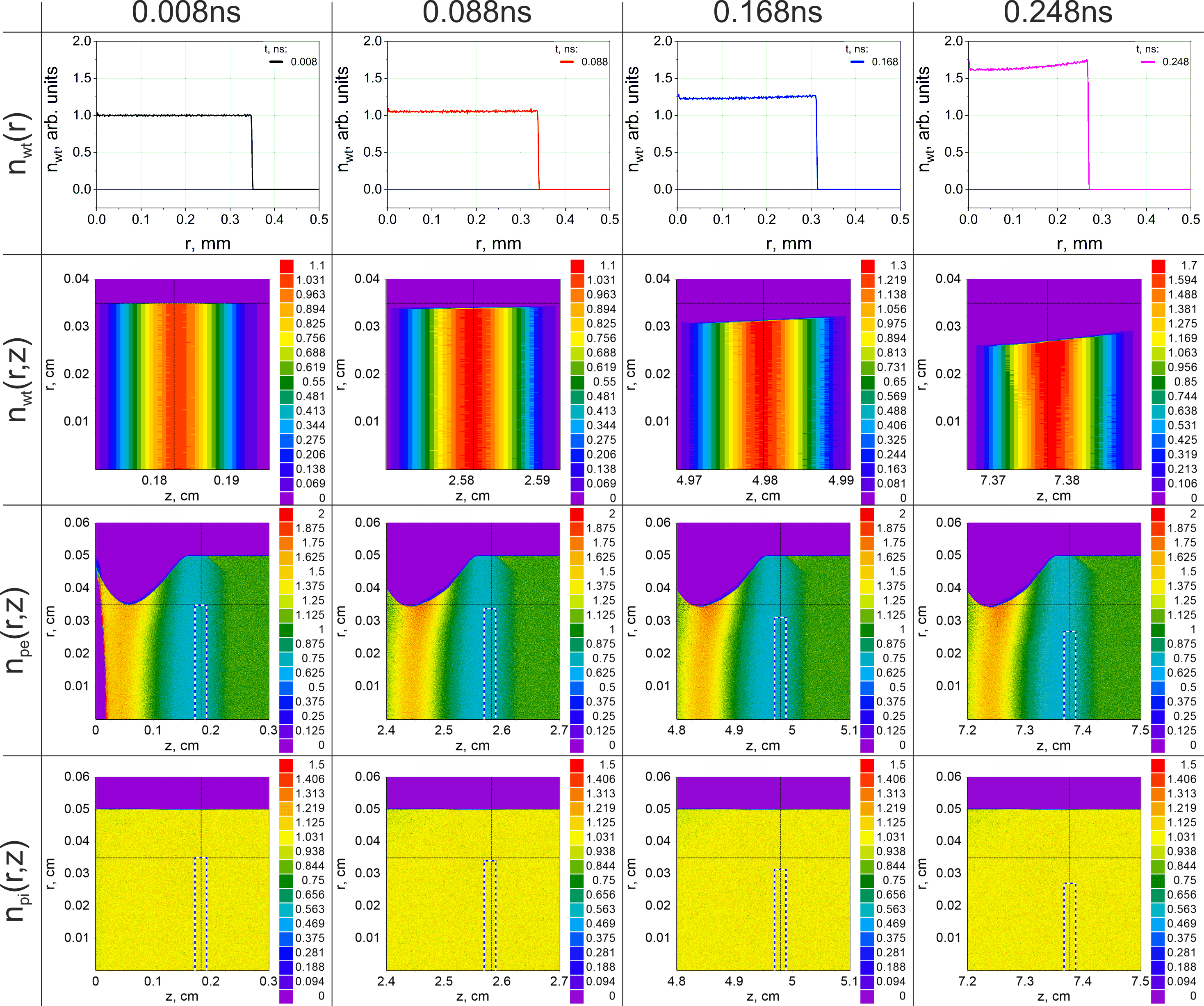}
  \caption{The density profile of the test electron bunch in the cross section passing through its middle $n_{wt}(r)$, and also color maps: the density of the test electron bunch $n_{wt}(r,z)$, the density of electrons $n_{pe}(r,z)$ and ions $n_{pi}(r,z)$ of the plasma for four moments of time: 0.008, 0.088, 0.168 and 0.248~ns. The plasma completely fills the transport channel, $r_{\rm v}=0$.}
  \label{Fig:02}
\end{figure}

\begin{figure}[!th]
  \centering
  \includegraphics[width=\textwidth]{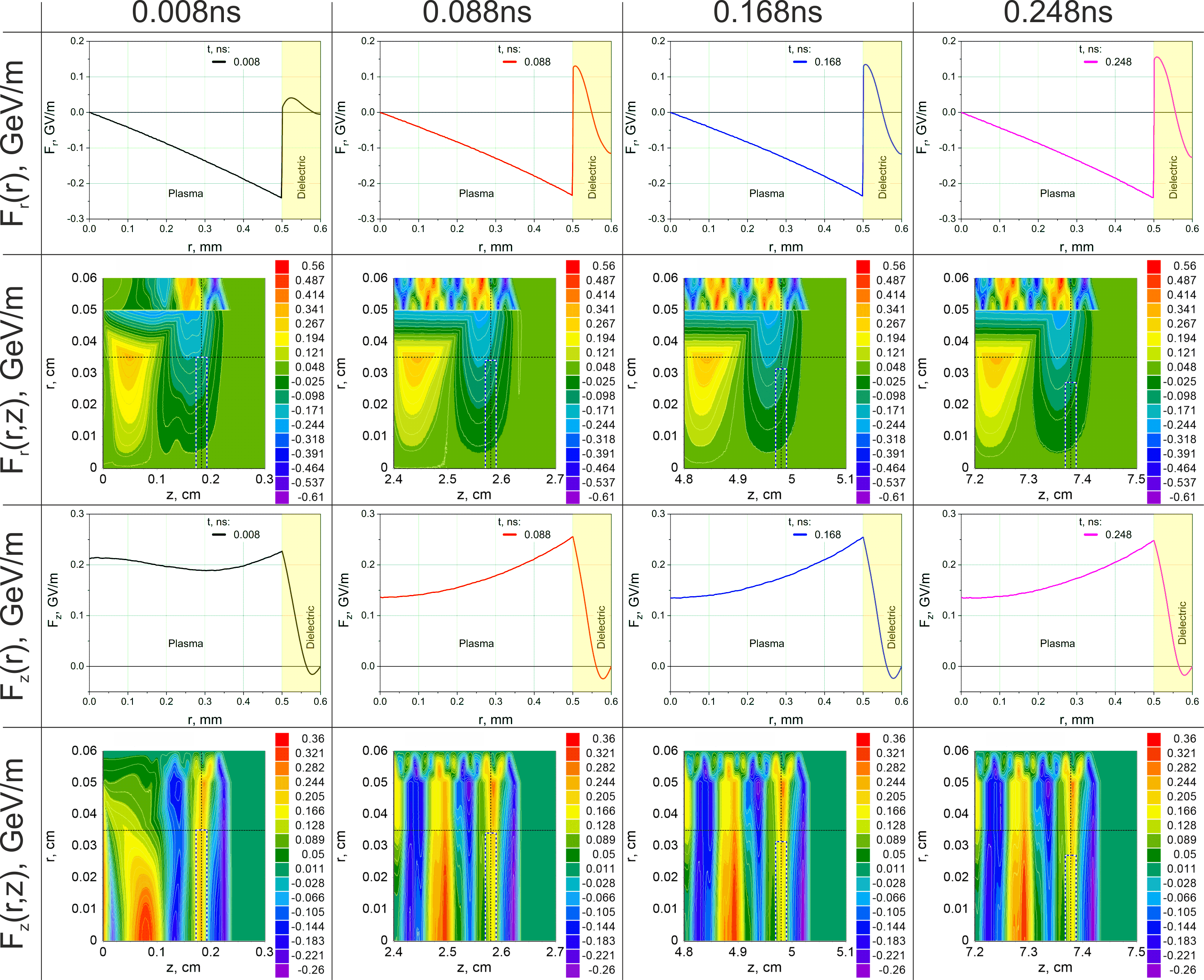}
  \caption{Profiles of the transverse $F_r(r)$ and longitudinal $F_z(r)$ forces acting on a test electron in a cross section passing through the middle of the test electron bunch, as well as color maps of the transverse $F_r(r,z)$ and longitudinal $F_z(r,z)$ forces acting on a test electron for the same moments of time and the same $r_{\rm v}$ as in Fig.~\ref{Fig:02}.}
  \label{Fig:03}
\end{figure}

\begin{figure}[!th]
  \centering
  \includegraphics[width=\textwidth]{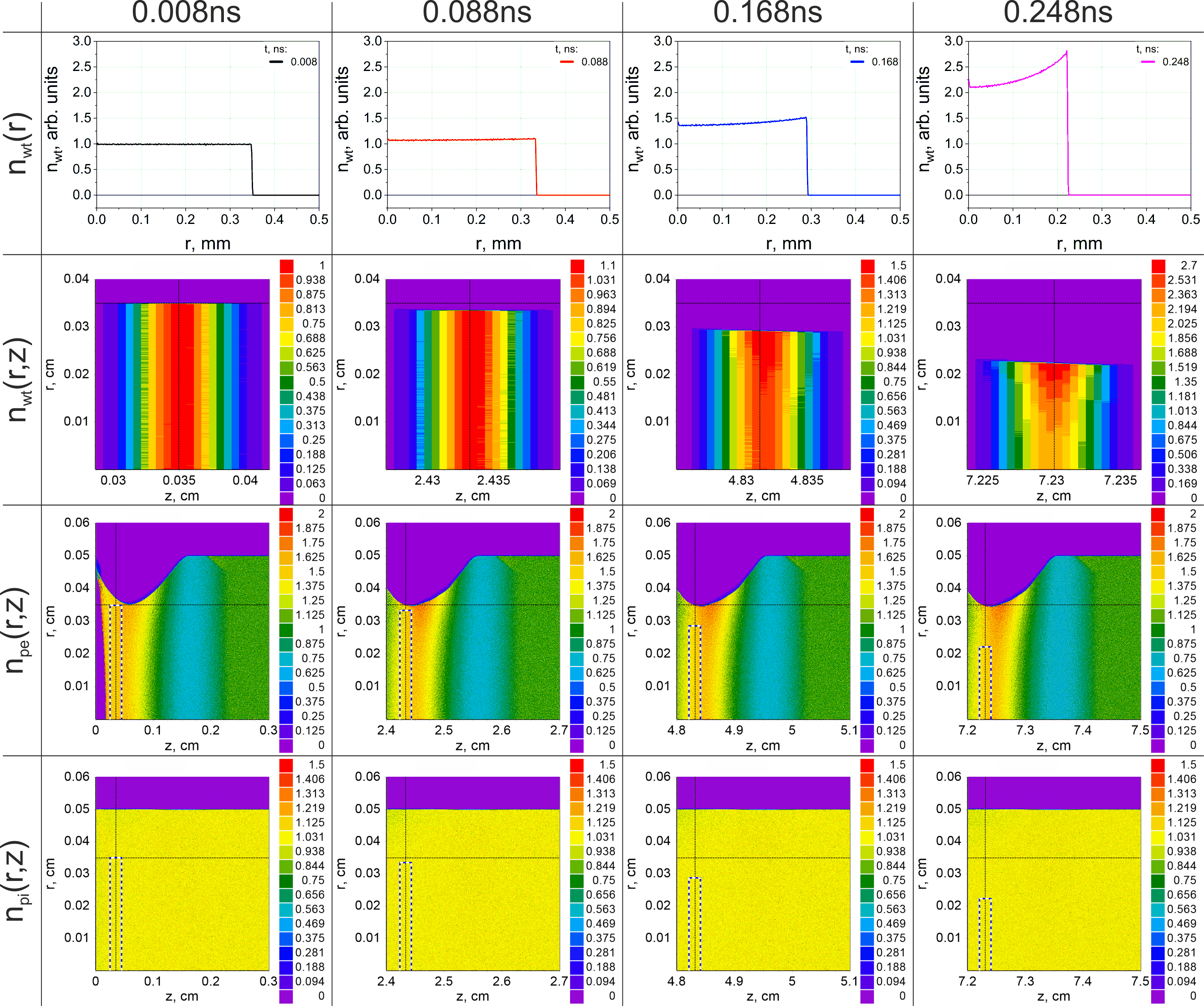}
  \caption{The same as in Fig.~\ref{Fig:02}, but for the case of a test positron bunch.}
  \label{Fig:04}
\end{figure}

\begin{figure}[!th]
  \centering
  \includegraphics[width=\textwidth]{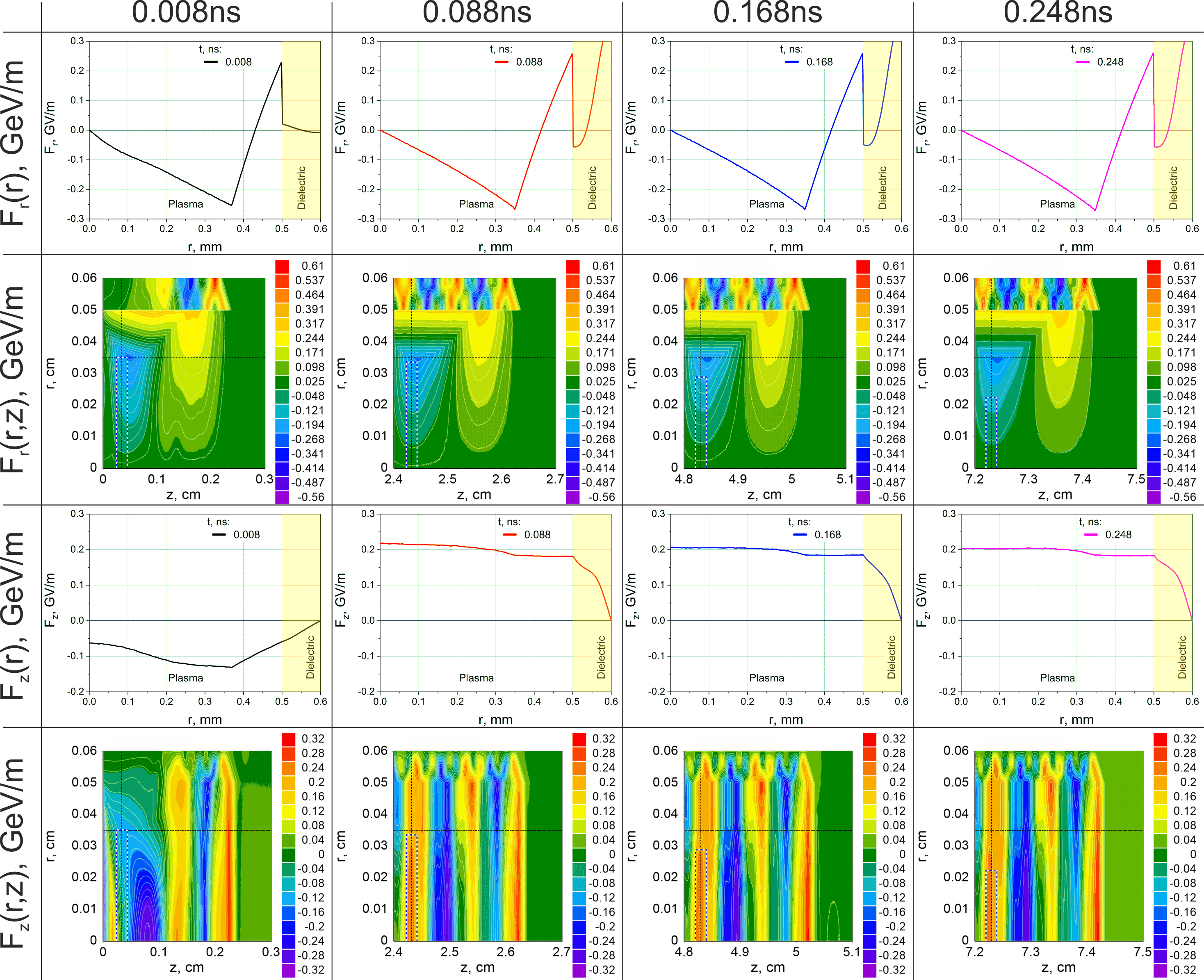}
  \caption{The same as in Fig.~\ref{Fig:03}, but for the forces acting on the test positron, for the case of a test positron bunch.}
  \label{Fig:05}
\end{figure}

The reason for the change in the dimensions of the test bunch and its acceleration when moving in the wake field created by the driver electron bunch is demonstrated in Figs.~\ref{Fig:02}-\ref{Fig:03} for the electron bunch, and in Figs.~\ref{Fig:04}-\ref{Fig:05} for the positron bunch. These figures correspond to the case of the complete absence of a vacuum channel in the plasma, $r_{\rm v}=0$.
Fig.~\ref{Fig:02} shows the density profile of the test electron bunch in the cross section passing through its middle $n_{wt}(r)$, as well as color maps of the density of the test electron bunch $n_{wt}(r,z)$, the density of electrons $n_{pe}(r,z)$ and ions $n_{pi}(r,z)$ of the plasma for four time moments of 0.008, 0.088, 0.168 and 0.248~ns. Fig.~\ref{Fig:03} shows the profiles of the transverse $F_r(r)$ and longitudinal $F_z(r)$ forces acting on the test electron in the cross section passing through the middle of the test bunch, as well as color maps of the transverse $F_r(r,z)$ and longitudinal $F_z(r,z)$ forces acting on the test electron for the same time moments as in Fig.~\ref{Fig:02}.

For clarity, the color maps include vertical and horizontal dotted lines showing the position of the test bunch middle in the longitudinal direction and its initial radius, respectively. The current position of the test bunch is marked on the graphs with a white and blue dotted line. As follows from the graphs $n_{pe}(r,z)$ in Fig.~\ref{Fig:02}, the wake field excited by the driver leads to the fact that the plasma electron density at the location of the test bunch decreases significantly (by about 40\%) compared to the initial value. At that, the plasma ion density $n_{pi}(r,z)$ does not have time to change while the test bunch is moving. This leads to the emergence of a negative (focusing) transverse force $F_r$ acting on the test electron. As follows from the graph $F_r(r)$ in Fig.~\ref{Fig:03}, this force changes almost linearly along the middle of the test bunch from the maximal negative value at the upper boundary of the drift chamber (at $r=a$) to zero on the system axis (at $r=0$).  As a consequence, the electrons of the test bunch move toward the system axis. In this case, the test bunch density (see top graphs $n_{wt}(r)$) increases almost uniformly in the transverse direction due to the displacement of the bunch electrons toward the axis. The transverse size of the bunch decreases and its focusing is observed.  Since the test bunch is in the region of positive values of the longitudinal force $F_z(r,z)$ acting on the test electron, which can be seen in the two lower rows of the graphs in Fig.~\ref{Fig:03}, the bunch accelerates as it moves in the drift chamber. From these same dependencies it follows too that the change in the acceleration rate for axial bunches is associated with the shift of the bunch in the phase of the accelerating wave.

Note that if we increase the test bunch size to the size of the drive bunch, i.e. when $r_w = R_b = 0.45$~mm, the focusing force $F_r$ acting on the test electron increases by approximately 1.35 times, and the force $F_z$ increases by 1.19 times. Therefore, with such parameters of the test electron bunch, both its focusing and acceleration will increase.

Fig.~\ref{Fig:04} and Fig.~\ref{Fig:05} are similar to Fig.~\ref{Fig:02} and Fig.~\ref{Fig:03}, but show the particle densities and forces acting on the test positron when test positron bunch injected into the system, respectively.
However focusing of an accelerated positron bunch is qualitatively different from focusing of an accelerated electron bunch. As follows from the graphs $n_{pe}(r,z)$ in Fig.~\ref{Fig:04}, the plasma electron density at the location of the test positron bunch is significantly higher (by about 40-50\%) compared to the initial value. These are the plasma electrons that were first pushed by the drive bunch to the periphery of the transport channel and then returned back to the axial region. This behavior of plasma electrons is similar to that given in the work~\cite{Diederichs_PRAB2019}. The plasma ion density $n_{pi}(r,z)$ changes very little during the injection delay of the test bunch. This leads to the emergence of a negative (directed toward the axis) transverse force acting on the test positron. As follows from the graph $F_r(r)$ in Fig.~\ref{Fig:05}, this force changes in the plane passing through the middle of the test bunch from the maximum negative value at $r\approx r_w$ to zero at $r=0$. As a result, the positrons of the test bunch are shifted downwards. According to the graph $F_r(r)$ in Fig.~\ref{Fig:05}, the force focusing the test positrons changes practically linearly and shifts all positrons of the bunch toward the axis. As a result, the transverse size of the bunch decreases, i.e., its focusing is observed. In accordance with Fig.~\ref{Fig:04}, the density of the positron bunch $n_{wt}(r)$ during its translational motion increases with time practically uniformly by the cross section. As follows from the dependencies $F_z(r)$ and $F_z(r,z)$ (two lower rows in Fig.~\ref{Fig:05}), focusing of the positron bunch is accompanied by its simultaneous acceleration not from the very beginning of its injection into the PDWA, but at $t>8$~ps. The start moment of the test bunch injection is chosen so as to obtain the optimal energy gain at the end of the accelerating structure. This optimization resulted in the positron bunch initially being in the region of negative values of the longitudinal force $F_z$.  During this period of time, the accelerating wave\footnote{Note that two waves are responsible for bunch transportation in PDWA - the accelerating wave, due to the presence of a dielectric, and the focusing wave due to the presence of plasma~\cite{Sot2014NIMA}} is not sufficiently formed due to boundary effects~\cite{Bal2001JETP,ONI2002PRE}.

Note that the chosen size of the positron bunch $r_{w} = 0.35$~mm is the most optimal from the point of view of its focusing. As follows from Fig.~\ref{Fig:05} (first row), with an increase in the bunch size compared to the optimal one up to 0.42~mm, a smaller focusing force will act on the peripheral positrons compared to the optimal one, and at $r_w > 0.42$~mm, the peripheral positrons will be under the action of a defocusing force. As for the bunch acceleration, due to the fact that the longitudinal force $F_z$ has a decreasing behavior with increasing radius $r$ (see Fig.~\ref{Fig:05}, third row, $t>8$~ps), an increase in the positron bunch size compared to the optimal one will weakly affect the change in its acceleration.

\begin{figure}[!th]
  \centering
  \includegraphics[width=0.5\textwidth]{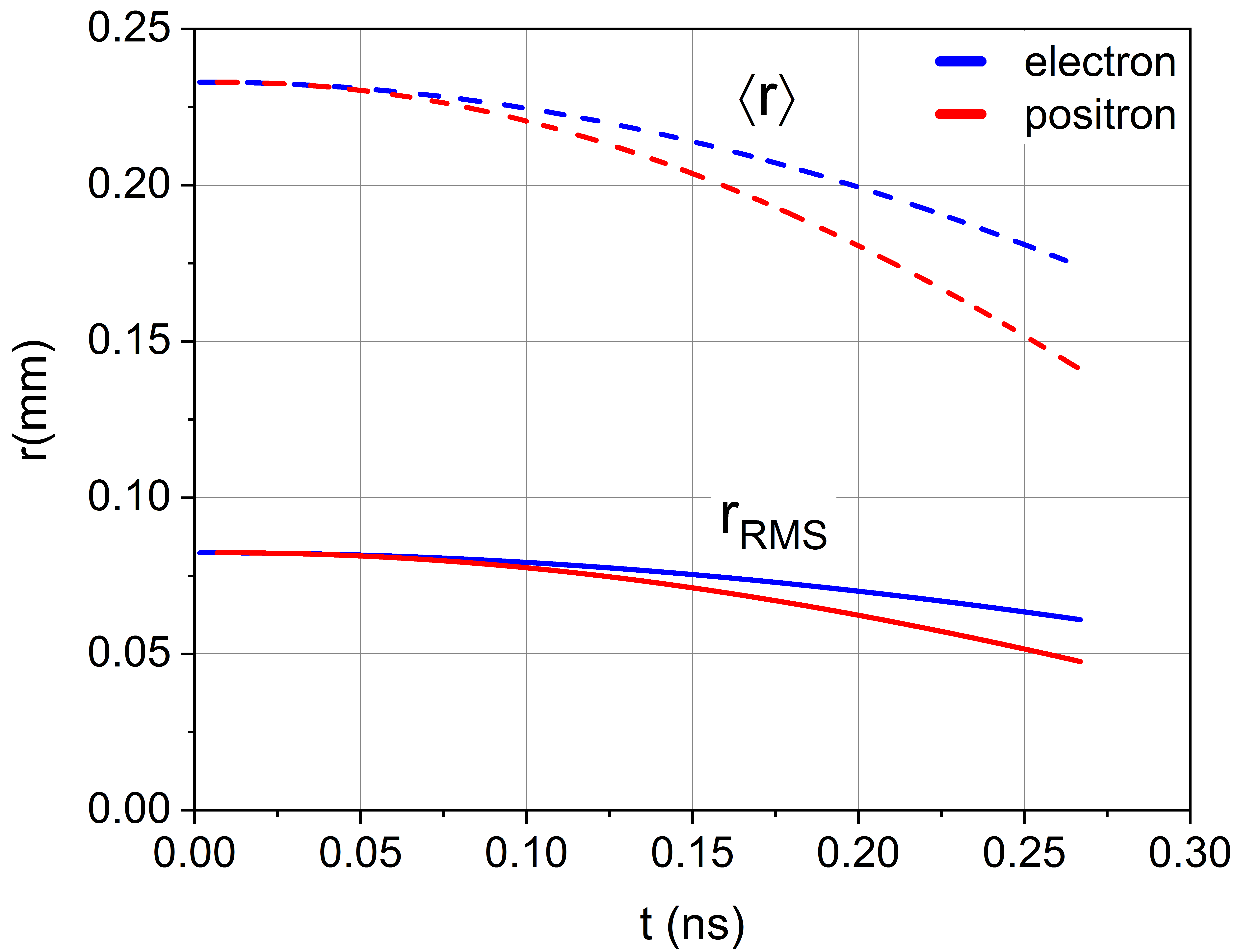}
  \caption{Evolution of the average particle radii $\langle r \rangle$ and the root-mean-square deviations of the particle radii from the average values $r_{RMS}$ of test electron and positron bunches when the drift channel is completely filled with plasma.}\label{Fig:06}
\end{figure}

\begin{figure}[!th]
  \centering
  \includegraphics[width=\textwidth]{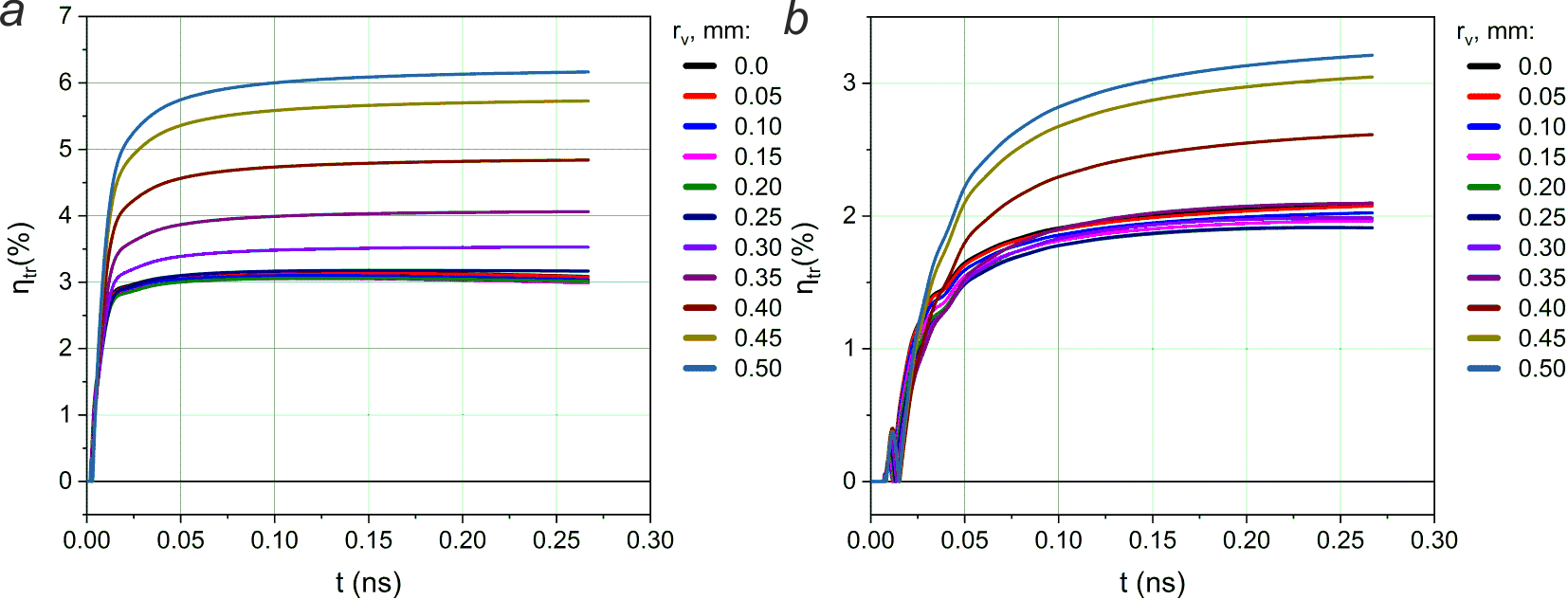}
  \caption{Efficiency of energy transfer $\eta_{tr}$  from the drive electron bunch to the test electron bunch (a) and positron bunch (b) depending on time for different radius of the vacuum channel $r_{\rm v}$. The time corresponds to the start of the driver electron bunch injection.}\label{Fig:07}
\end{figure}

\begin{figure}[!th]
  \centering
  \includegraphics[width=\textwidth]{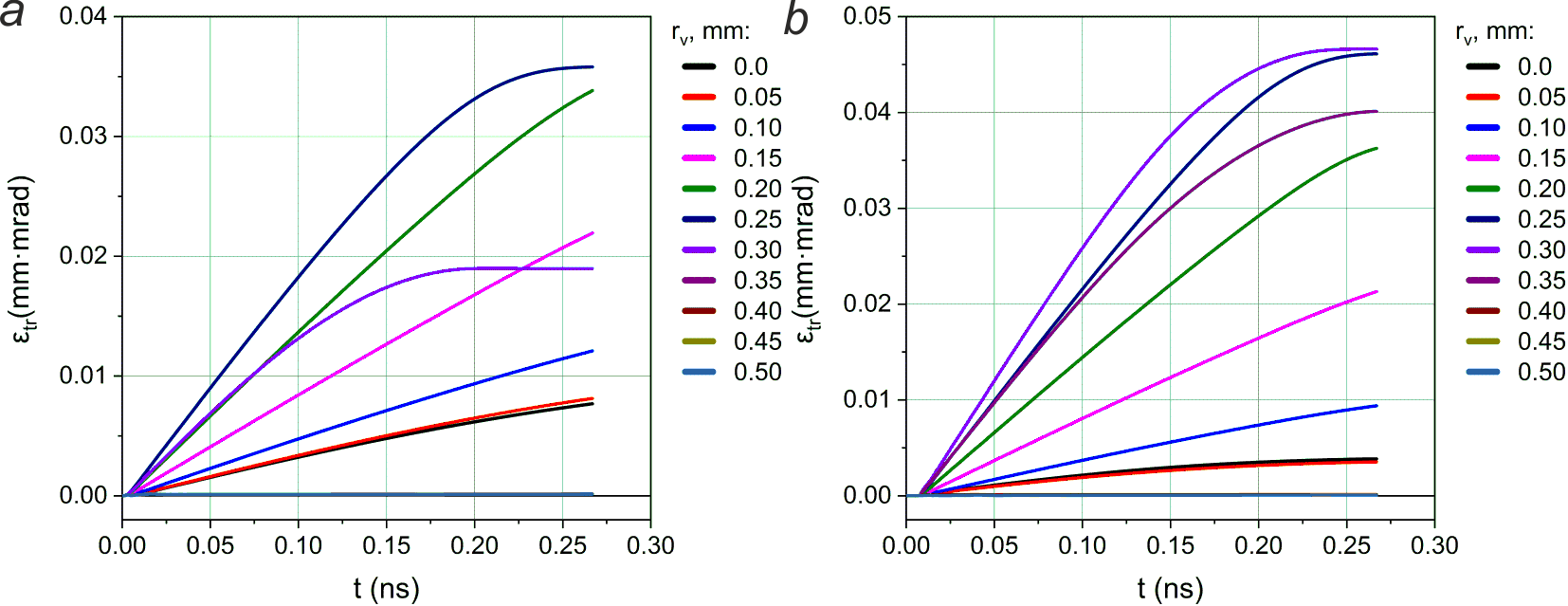}
  \caption{Evolution of the emittance of the test bunch of electrons (a) and positrons (b) for different radius of the vacuum channel $r_{\rm v}$.}\label{Fig:08}
\end{figure}

\begin{figure}[!th]
  \centering
  \includegraphics[width=\textwidth]{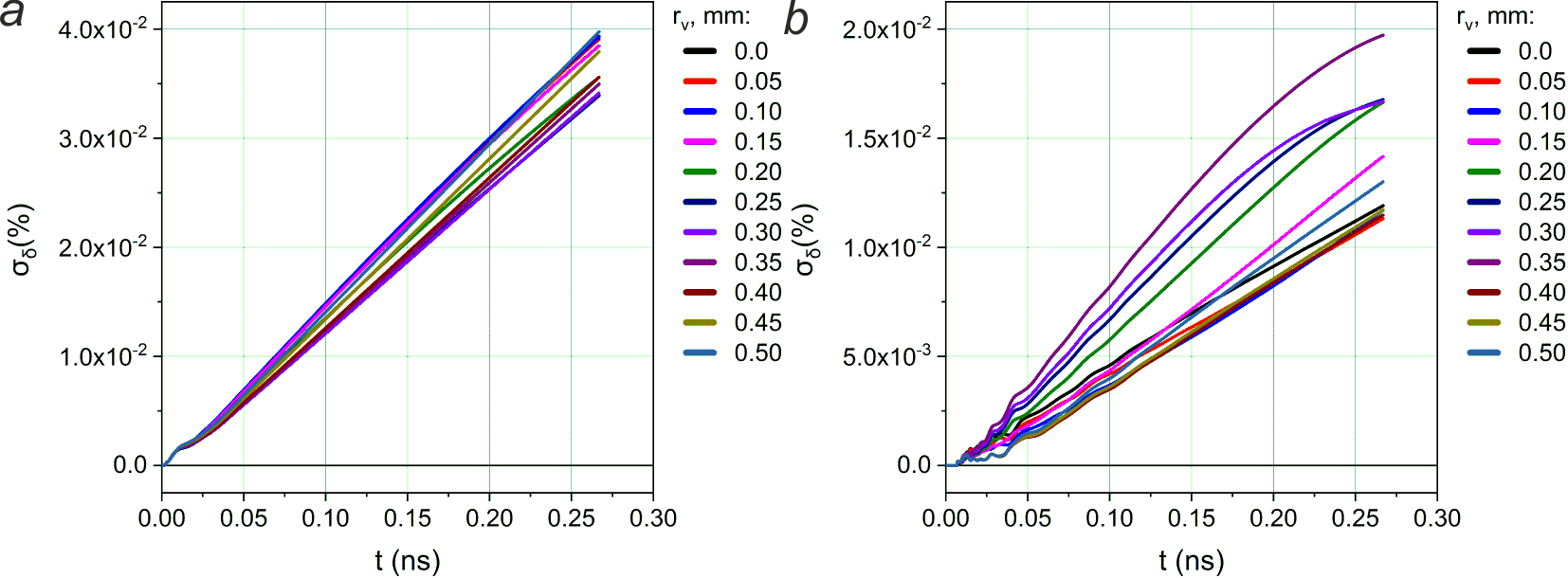}
  \caption{Energy spread of the test bunch of electrons (a) and positrons (b) as a function of time for different radius of the  vacuum channel $r_{\rm v}$.}\label{Fig:09}
\end{figure}

Let us analyze the characteristics of the accelerated electron and positron bunches. Figs.~\ref{Fig:06}--\ref{Fig:09} shows the root-mean-square deviation of the radii, the efficiency of energy transfer to the test bunch from the driver one, the emittances of the test bunches and their energy spread.

To demonstrate the focusing of accelerated bunches in Fig.~\ref{Fig:01} we used $R_{max}$ -- the maximum radius of a particle in a bunch. Such characteristic qualitatively confirms the transverse dynamics of bunches, but for a quantitative characteristic it is necessary to use the averaged radii of all particles in the bunch. This is especially relevant in the case of using macroparticles of different charges in numerical simulation. The dynamics of the average particle radii change $\langle r \rangle=\sum\limits_i r_i\cdot q_i/Q$, where $r_i$, $q_i$ are the radius and charge of the macroparticle, $Q$ is the total charge of the bunch, and the root-mean-square deviations of the particle radii from the average values $r_{RMS}=\sqrt{\sum\limits_i (r_i-\langle r \rangle)^2\cdot q_i/Q}$ of the test electron and positron bunches when the drift channel is completely filled with plasma are demonstrated in Fig.~\ref{Fig:06}. It is evident that the decrease in the average radius $\langle r\rangle$ of the test positron bunch is greater than that of the electron bunch, which coincides with the qualitative behavior of $R_{max}$ (Fig.~\ref{Fig:01}) and is associated with a stronger focusing of the positron bunch. Also, the transverse spatial spread of particles of the positron and electron bunches $r_{RMS}$ decreases when moving along the accelerator and is smaller for the positron bunch.

To calculate the efficiency of energy transfer from the drive bunch to the witness bunch as the ratio of the acquired energy of the test bunch to the energy loss of the drive bunch, the emittances of the test bunches and their energy spread we will also use the value averaged over macroparticles. The expressions for calculations of these values accounting a macroparticle weight are given in ref.~\cite{Markov_2022}.

 Figure~\ref{Fig:07} shows the behaviour of efficiency $\eta_{tr}$ of energy transfer from the drive electron bunch to the accelerated bunches. The maximum efficiency for the electron test bunch was about 6.16\% and for the positron one 3.21\% and was achieved in the complete absence of plasma in the drift channel. The minimum efficiency was achieved at $r_{\rm v}$, which ensures the best focusing of the test bunch, as can be seen in Fig.~\ref{Fig:07}a: for the electron test bunch it is $\approx3\%$; for the positron test bunch $\approx1.91\%$ (Fig.~\ref{Fig:07}b). In the presence of a vacuum channel, the efficiency is lower, which is due to the lower value of the accelerating field.

In Fig.~\ref{Fig:08} the change in emittance~\cite{Floetman_PhysRevSTAB.6.034202} of electron and positron bunches during acceleration in the PDWA is shown. As follows from Fig.~\ref{Fig:08}, the emittance of the test beam increases with acceleration in the PDWA, but after passing $\approx 7$~cm it increases to acceptable values. The lowest emittance value of $1.29\cdot10^{-4}$~mm$\cdot$mrad for electron test beam and $7.85\cdot10^{-5}$~mm$\cdot$mrad for positron beam is realized for the case when the drift channel is completely free of plasma. In case of incomplete filling, the emittances are higher for both the test electron and positron bunches. The highest emittance value is obtained in the case of the best focusing of the test bunch. For the test electron bunch it is $3.57\cdot10^{-2}$~mm$\cdot$mrad, and $4.66\cdot10^{-2}$~mm$\cdot$mrad for the positron bunch. It can be seen that even for this nonoptimal case there is already a tendency to reach saturation.

Figure~\ref{Fig:09} shows the change in the energy spread of the electron (a) and positron (b) test bunches during acceleration along the PWFA. For the electron test bunch (Fig.~\ref{Fig:09}a) $\sigma_\delta$ is practically a linear dependence on time $t$.
The smallest spread of the electron bunch energy is realized at $r_{\rm v}$ providing the best bunch focusing and is equal to 0.034\%. The largest energy spread is realized when the drift channel is completely filled with plasma and is equal to 0.039\%. For the positron bunch (Fig.~\ref{Fig:09}b) the smallest energy spread $(\sigma_\delta = 0.011\%)$ is realized at $r_{\rm v}=0.05$~mm, i.e. when the drift channel is almost completely filled with plasma. The largest energy spread of 0.022\% is realized at $r_{\rm v}=0.375$~mm, i.e. when the radius of the vacuum tube is larger than the radius of the test bunch. With a further increase in $r_{\rm v}$ the energy spread of the positron bunch drops sharply to a minimum value. These values of the energy spread are significantly lower than the requirements for collider applications~\cite{Shiltsev_2012,Diederichs_PRAB2020, Svensson_NIMA_2022,Liu2024PRL}.

\section{BBU instability in PDWA}\label{sec:4}
BBU instability of the drive bunch in PDWA will be considered in the general statement of the section~\ref{sec:2}. The plasma completely fills the channel of the dielectric tube. In contrast to the studies of the previous section, drive bunch is injected with a transverse offset $R_{off}$ with respect of the accelerating structure axis.

First, we give the analytical results. In the linear approximation, having solved the system of Maxwell's equations, we obtain expressions for the components of the excited wakefield. To analyze the BBU, we will be interested only in the transverse wakefield $W_r=E_{r}-\beta H_{\phi}$. The analytical expressions for the force acting on the particles located at the upper ($r = R_{off} + R_b$) and lower ($r = R_{off} - R_b$) boundaries of the drive bunch have the following form~\cite{GALAYDYCH2022166766}:
\begin{equation}\label{eq:01}
\begin{split}
&W_r(r = {R_{off}} + {R_b},\varphi ,\xi ) = \frac{4Q_b}{R_b}\sum\limits_{m =  - \infty }^{ + \infty } {{e^{im\varphi }} \times } \\
&\bigg(\frac{{{I_m}({k_p}{R_{off}}){I_1}({k_p}{R_b})}}{{{I_m}({k_p}a)}}{\Delta '_m}({k_p}a,{k_p}({R_{off}} + {R_b}))\Psi _ \bot ^{(p)}(\xi ) + \\
&\sum\limits_{s = 1} {\frac{{2{v^2}{D_2}({\omega _s})}}{{a\omega _s^3D'({\omega _s})}}\frac{{{I_m}({\kappa _{ps}}{R_{off}}){I_1}({\kappa _{ps}}{R_b})}}{{I_m^2({\kappa _{ps}}a)}}}  \times \\
&{I'_m}({\kappa _{ps}}({R_{off}} + {R_b}))\Psi _ \bot ^{(s)}(\xi )\bigg),
\end{split}
\end{equation}
\begin{equation}\label{eq:02}
\begin{split}
&W_r(r = {R_{off}} - {R_b},\varphi ,\xi ) = \frac{4Q_b}{R_b}\sum\limits_{m =  - \infty }^{ + \infty } {{e^{im\varphi }}}  \times \\
&\bigg(\frac{{{I'_m}({k_p}({R_{off}} - {R_b})){I_1}({k_p}{R_b})}}{{{I_m}({k_p}a)}}{\Delta _m}({k_p}a,{k_p}{R_{off}})\Psi _ \bot ^{(p)}(\xi ) + \\
&\sum\limits_{s = 1} {\frac{{2{v^2}{D_2}({\omega _s})}}{{a\omega _s^3D'({\omega _s})}}\frac{{{I_m}({\kappa _{ps}}{R_{off}}){I_1}({\kappa _{ps}}{R_b})}}{{I_m^2({\kappa _{ps}}a)}}}  \times \\
&{I'_m}({\kappa _{ps}}({R_{off}} - {R_b}))\Psi _ \bot ^{(s)}(\xi )\bigg),
\end{split}
\end{equation}
where the following notations are used: $\xi = t - z/v$, $\kappa _{p}^2 = (\omega/v)^2(1 - {\beta ^2}{\varepsilon _p}({\omega}))$, ${\Delta _m}(x,y) = {I_m}(x){K_m}(y) - {K_m}(x){I_m}(y)$, $I_m$ and $K_m$ are the modified Bessel and Macdonald functions of the $m^{th}$ order, ${\varepsilon _p}(\omega ) = 1 - \omega _p^2/{\omega ^2}$ is the plasma permittivity, $\omega_p$ is the plasma frequency, $k_p=\omega_p/v$ is the plasma wavenumber, $\omega_s$ are the frequencies of the plasma-dielectric waveguide eigenmodes, which are in Cherenkov resonance with the drive bunch, and are defined by the roots of the equation $D(\omega_s)=0$, ${\kappa _{ps}} = {\kappa _p}({\omega _s})$, $D'({\omega _s})=dD({\omega_s})/d\omega_s$. Resonant with the drive bunch eigenfrequencies of azimuthally homogeneous modes ($m=0$) are determined from the solution of the dispersion equation
\begin{equation}\label{eq:03}
\begin{split}
D(\omega) \equiv\frac{{{\varepsilon _p}(\omega )}}{{{\kappa _p}}}\frac{{{{I'}_0}({\kappa _p}a)}}{{{I_0}({\kappa _p}a)}} + \frac{{{\varepsilon _d}}}{{{\kappa _d}}}\frac{{{{F'}_0}({\kappa _d}a,{\kappa _d}b)}}{{{F_0}({\kappa _d}a,{\kappa _d}b)}} = 0.
\end{split}
\end{equation}
The resonant with the drive bunch eigenfrequencies of azimuthally inhomogeneous modes ($m\neq0$) are determined from the solution of the dispersion equation
\begin{equation}\label{eq:04}
\begin{split}
D(\omega) &\equiv D_1(\omega)D_2(\omega) - \frac{\beta^2m^2(\omega/v)^4}{a^2\kappa_p^4\kappa_d^4}(\varepsilon_d-\varepsilon_p)^2 = 0,\\
D_1(\omega) &= \frac{\varepsilon_p}{\kappa_p}\frac{I^\prime_m(\kappa_p a)}{I_m(\kappa_p a)} + \frac{\varepsilon_d}{\kappa_d}\frac{F^\prime_m(\kappa_d a, \kappa_d b)}{F_m(\kappa_d a, \kappa_d b)},\\
D_2(\omega) &= \frac{1}{\kappa_p}\frac{I^\prime_m(\kappa_p a)}{I_m(\kappa_p a)} + \frac{1}{\kappa_d}\frac{\Phi^\prime_m(\kappa_d a, \kappa_d b)}{\Phi_m(\kappa_d a, \kappa_d b)},
\end{split}
\end{equation}
where $\kappa_d^2=(\omega/v)^2(\beta^2\varepsilon_d-1)$, $\varepsilon_d$ is the dielectric permittivity, ${F_m}(x,y) = {( - 1)^m}({J_m}(x){Y_m}(y) - {Y_m}(x){J_m}(y))$, $J_m$ and $Y_m$ are the Bessel and Weber functions of the order $m^{th}$, ${F'_m}(x,y) = {( - 1)^m}({J'_m}(x){Y_m}(y) - {Y'_m}(x){J_m}(y))$, ${\Phi _m}(x,y) = {J_m}(x){Y'_m}(y) - {Y_m}(x){J'_m}(y)$, ${\Phi '_m}(x,y) = {J'_m}(x){Y'_m}(y) - {Y'_m}(x){J'_m}(y)$.
For the bunch with homogeneous longitudinal charge distribution the axial structure of the excited transverse wakefield $\Psi_{\perp}^{p,s}(\xi)$ have the following form
\begin{equation}\label{eq:05}
\begin{split}
\Psi_{\perp}^{p,s}(\xi) = \left[\theta (\xi)(1 - \cos {\omega _{p,s}}\xi)-\theta (\xi - L_b/v)(1 - \cos {\omega _{p,s}}(\xi - L_b/v))\right]/L_b,
\end{split}
\end{equation}
where $\theta$ is the Heaviside function, $L_b$ is the bunch length.

The analytical expressions~(\ref{eq:01})-~(\ref{eq:02}) demonstrate the fact that the total excited wakefield consists of the wakefield of the waveguide eigenwaves, and the plasma wakefield.

For numerical analysis of the expressions for the transverse wakefield, we will use the following drive bunch parameters: $L_b=0.5$ mm, $R_b=0.23$ mm, $R_{off}=0.24$ mm, the other variables entering in the expressions (\ref{eq:01})-(\ref{eq:05}) are the same as in Table~\ref{Tabl:01}.

\begin{figure}[!th]
  \centering
  \includegraphics[width=0.75\textwidth]{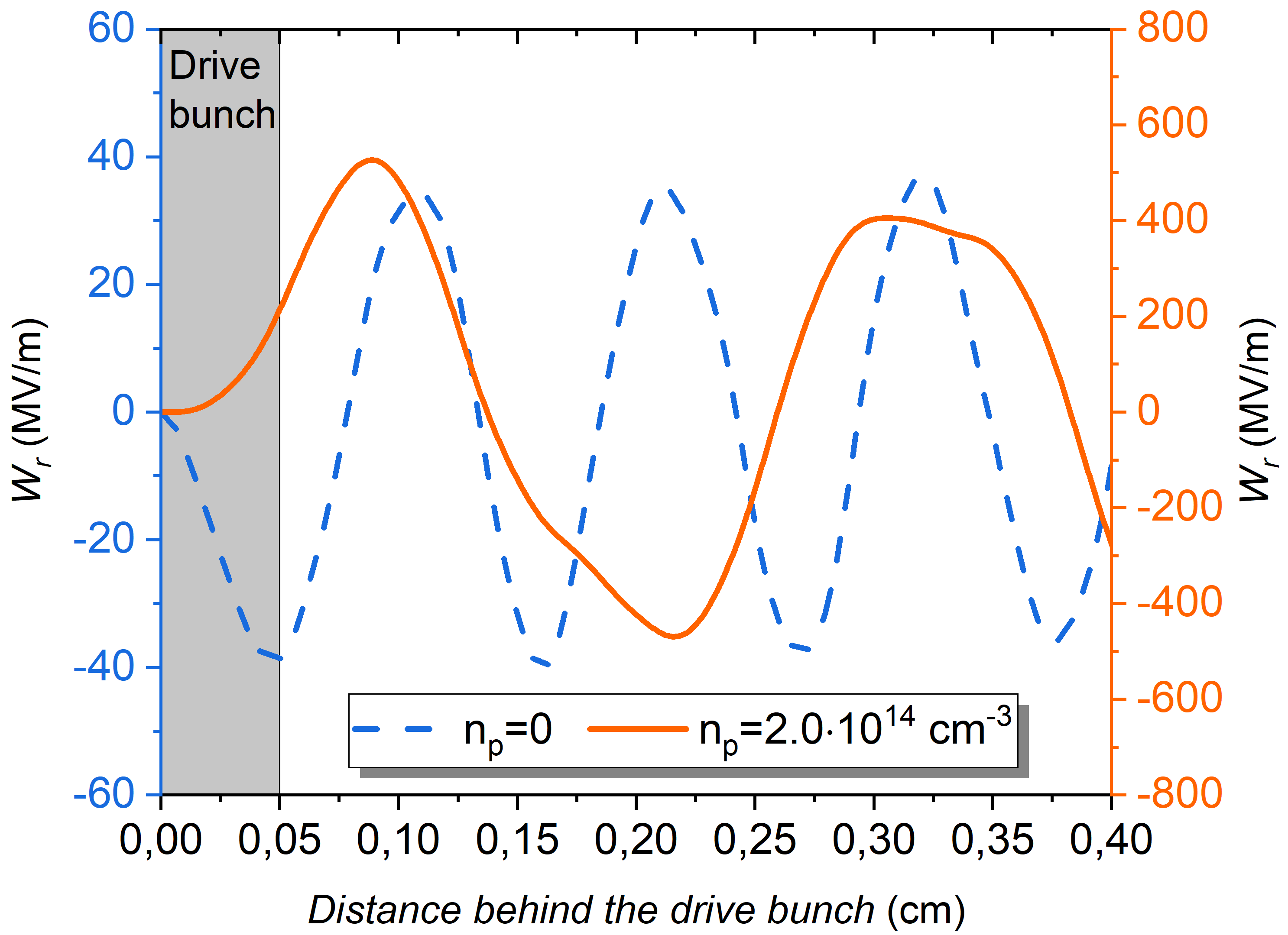}
  \caption{Longitudinal distribution of the radial wakefield $W_r$ excited by the drive electron bunch for the cases of absence and presence of plasma filling calculated at $r=R_b+R_{off}$. The bunch moves from right to left parallel to the waveguide axis with constant velocity.}\label{Fig:11}
\end{figure}

The transverse wakefield excited in the accelerating structure by the drive bunch determines the radial dynamics of both the drive bunch and the test bunch, which can be accelerated in this field. Fig.~\ref{Fig:11} shows the longitudinal distributions of the transverse wakefield amplitude for the cases of absence and presence of plasma filling, calculated on the surface of the drive bunch (at $r=R_b+R_{off}$).
In the case of a vacuum channel for charged particles, the transverse wakefield in the region of the drive bunch is defocusing. The amplitude of this field grows from the head to the tail of the bunch. This leads to the fact that the bunch starts to deflect as a whole to the dielectric (the loss of charge of the bunch starts at its tail). In the case of a channel for charged particles filled with plasma, the transverse wake field is focusing. In this case, in comparison with the case of a vacuum channel, not only the sign of the field amplitude in the region of the drive bunch changes, but also its characteristic spatial scale behind the bunch.
It is seen that the amplitude profile of the transverse wake field in the case of the presence of plasma has a more multi-wavelength structure compared to the case of the absence of plasma. This, in turn, may indicate that more modes take part in its formation.

\begin{figure}[!th]
  \centering
  \includegraphics[width=0.75\textwidth]{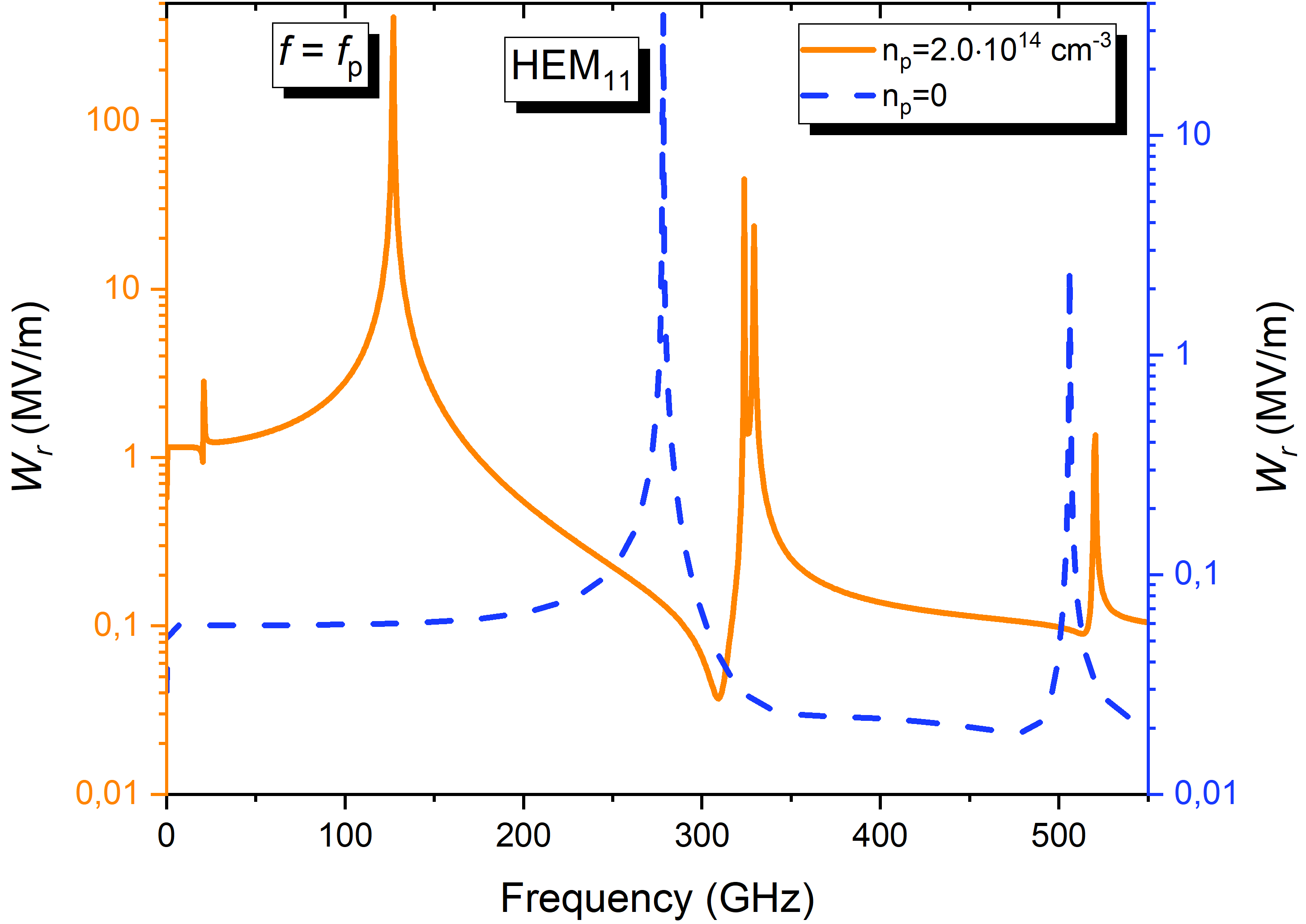}
  \caption{Spectra of the radial wake fields $W_r$, excited by the drive bunch, for the cases of absence and presence of plasma filling, calculated at $r = R_b + R_{off}$.}\label{Fig:12}
\end{figure}

In addition to analyzing the longitudinal amplitude distribution of the transverse wakefield, its spectral analysis was also performed. The frequency composition of the excited transverse wakefield in the charged particles transport channel is presented in Fig.~\ref{Fig:12} for the cases of absence and presence of plasma.
In the absence of plasma, the maximum in the spectrum of the transverse wakefield corresponds to the frequency of the dipole mode ($HEM_{11}$). Significant excitation of this parasitic mode leads to an undesirable deflection of the drive bunch as a whole at its injection with initial offset. In the presence of plasma in the charged particles transport channel, the picture of the spectrum of the transverse wake field changes significantly. Namely, the filling of the vacuum channel with plasma leads to a change in the frequency composition of the excited field. The main difference in this case is that the maximum in the spectrum of the transverse wakefield corresponds to the plasma wave frequency (127.28 GHz), which is absent in the case of no plasma. In addition, the spectrum contains the frequency of the working mode of the accelerating structure (329.29 GHz), higher-order modes (323.71 GHz, 520.11 GHz), and the resonant frequency of the surface wave (20.87 GHz), which propagates along the plasma-dielectric boundary and has the asymptote $\omega(k_z\rightarrow\infty)=\omega_p/\sqrt{1+\varepsilon_d}$. The frequencies of the higher-order modes, resonant with the bunch, as numerical studies have shown, are increased due to the presence of plasma. But the contribution of the above modes, compared to the contribution of the plasma wave, is not determinant in the transverse wakefield amplitude.

\begin{figure}[!th]
  \centering
  \includegraphics[width=0.49\textwidth]{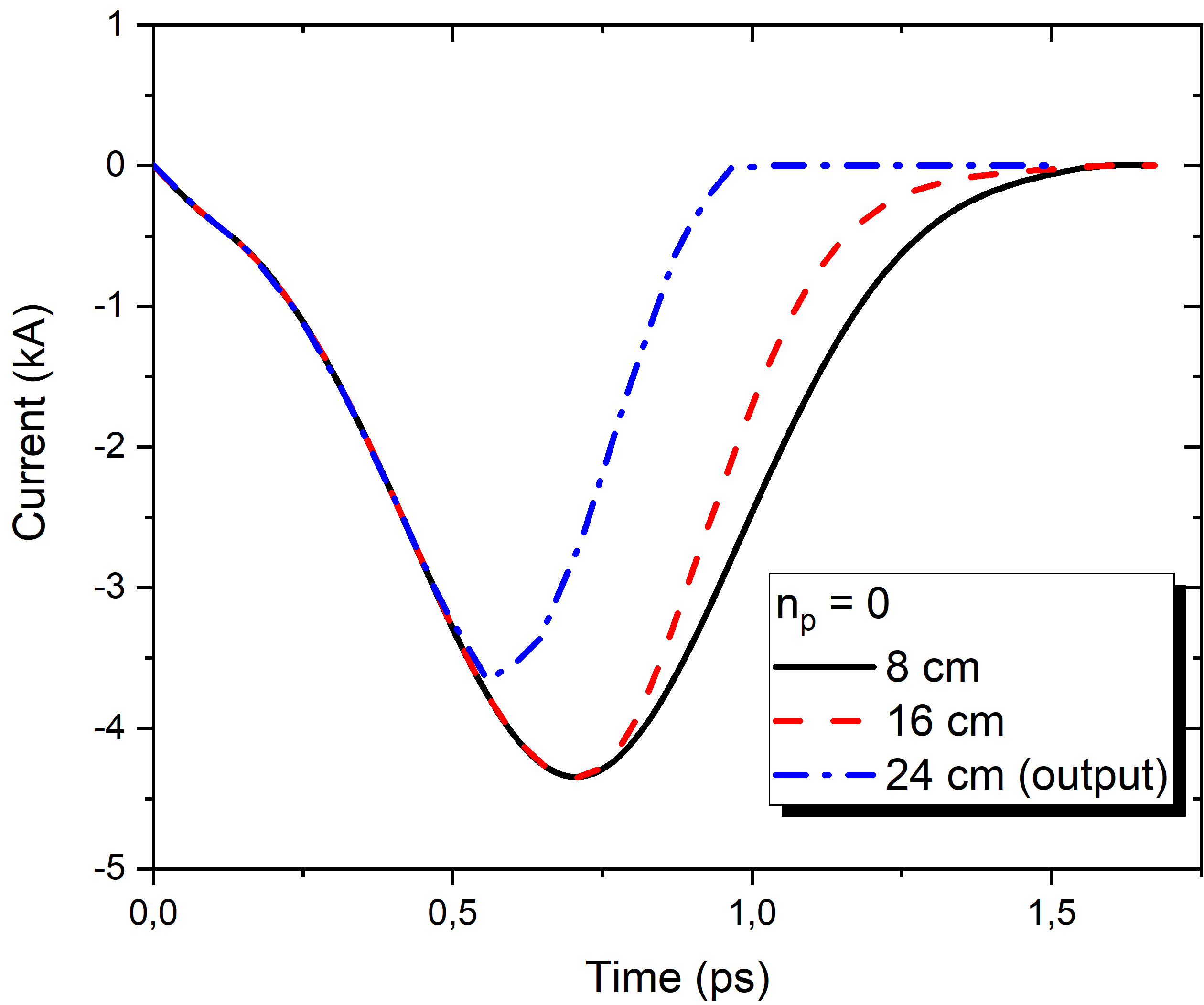}
  \includegraphics[width=0.49\textwidth]{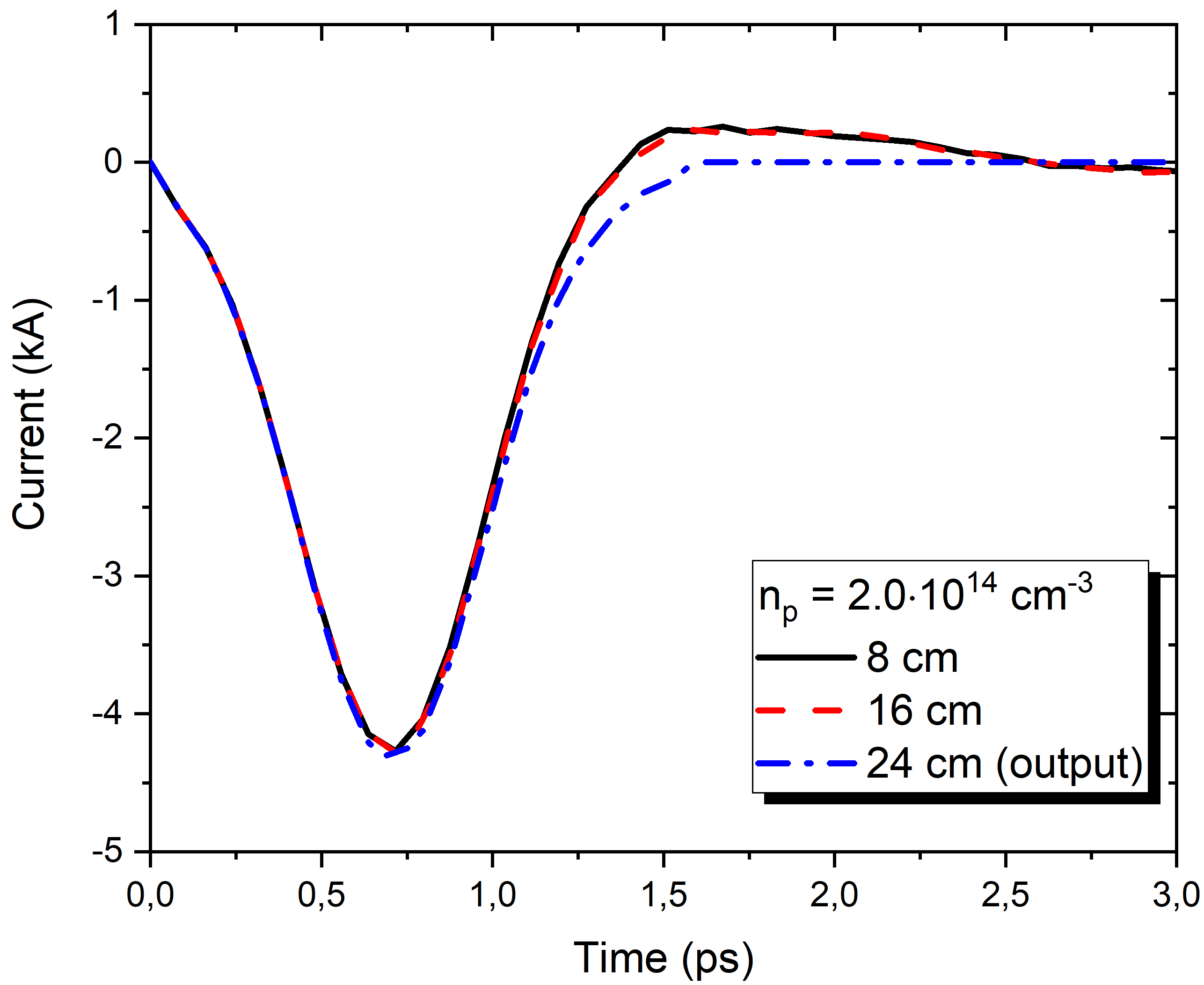}
  \caption{A current through a cross section of the transport channel of the accelerating structure as a function of time for longitudinal coordinates 8 cm, 16 cm and 24 cm (structure output): (left) $n_p = 0$, (right) $n_p = 2.0\cdot 10^{14}\,cm^{-3}$. Maximum bunch current in the simulations is 4.5 kA.}\label{Fig:13}
\end{figure}

Besides analytical analysis of the wakefield excitation in the linear appro\-ximation, PIC simulations were performed. The main goal of the simulations was to analyze the dynamics of the off-axis drive bunch. To demonstrate more clearly the effect of filling the charged particles channel with plasma, the length of the accelerating structure was chosen to be 24 cm and the transverse size of the bunch was chosen to be 0.12 mm. To analyze the BBU instability, the current through the cross section of the charged particles transport channel was chosen as a numerical characteristic. To demonstrate the dynamics of the drive bunch current dynamics, the current through the cross section of the waveguide was analyzed for the cases of absence and presence of plasma filling for three values of the longitudinal coordinate, namely: 8 cm, 16 cm, and 24 cm (at the exit of the waveguide). These results are presented in Fig.~\ref{Fig:13}.
This figure shows that in the absence of plasma, the maximum current of the drive bunch gradually decreases, and its duration decreases due to the deposition of bunch particles on the dielectric. In the presence of plasma, the value of the maximum current of the drive bunch practically does not change as it moves along the accelerating structure. In this case, there is no deposition of bunch particles on the dielectric surface, but due to the radial refocusing of the bunch in its tail, there is a gradual increase in its transverse size. The absence of reduction of the maximum value of the bunch current, in the presence of partial losses of its charge, occurs due to radial refocusing by the electric field of the plasma wave. It is also necessary to note the presence of a current with opposite sign (compared to the bunch current) at the end of the current pulse through the cross sections of the waveguide. This is the current of plasma electrons trapped by the excited wake field, the amplitude of which oscillates with the frequency of the eigenwave resonant with the bunch, and decreases with time as the bunch moves from the given cross section to the exit of the accelerating structure.

\section{Conclusions}
\label{sec:Conclusion}
In our studies, the main focus was attend on studying the characteristics of accelerated positron and electron bunches by numerical PIC modeling of the wake field excited by the drive electron bunch and the self-consistent dynamics of charged particles in the plasma-dielectric slow-wave structure of the THz range for a model of a homogeneous plasma filling the drift region in a waveguide.
For the most adequate comparison of the characteristics, we chose the same sizes of the test electron and positron bunches and the parameters of the PWFA. Only the injection moments of the electron and positron bunches were differed.
We show that the presence of plasma in the system leads to focusing of the accelerated test bunch of positrons or electrons, and the presence of a paraxial vacuum channel increases this focusing.
Different mechanisms are responsible for focusing electron and positron bunches. The plasma ions remaining after the plasma electrons are pushed to the periphery of the transport channel by the driver electron bunch are responsible for focusing the accelerated electron bunch. The plasma electrons returning from the periphery to the location of the positron bunch are responsible for focusing the test positron bunch.
In the absence of plasma, focusing of the test bunch is not observed, but in this case the energy gain of the bunch particles is the greatest. The characteristics of accelerated positron and electron bunches were investigated: transverse root-mean-square size, emittance, energy spread and efficiency of energy transfer from the driver electron bunch to the test one.

A comparison, using analytical theory and PIC simulation, of the off-axis drive bunch transverse dynamics for the cases of the absence and presence of plasma in the channel for charged particles showed, that this transverse dynamics has significant difference. Namely, this comparison showed that the BBU instability of the drive bunch does not develop according to the classical scenario, when there is a gradual deflection of the bunch particles in its tail. Instead, there is a focusing of the tail part of the drive bunch, and due to this significant reduction of its current does not occur. The analysis also showed that in the presence of plasma, the deposition of the bunch particles (due to refocusing) can start much later or not at all, depending on the length of the accelerating structure and the initial offset. The linear and quasi-linear numerical analysis allows us to conclude that the presence of plasma in the channel for charged particles leads to the mitigation of BBU instability.

\section*{Acknowledgments}
The study is supported by the National Research Foundation of Ukraine under the program “Excellent science in Ukraine” (project \# 2023.03/0182).

\bibliographystyle{elsarticle-num}
\bibliography{PDWA_NIMA}

\end{document}